\documentclass[sigconf]{acmart}
\acmConference[ISSTA 2023]{ACM SIGSOFT International Symposium on Software Testing and Analysis}{17-21 July, 2023}{Seattle, USA}
\AtBeginDocument{%
  \providecommand\BibTeX{{%
    \normalfont B\kern-0.5em{\scshape i\kern-0.25em b}\kern-0.8em\TeX}}}

\setcopyright{acmcopyright}
\copyrightyear{2018}
\acmYear{2018}
\acmDOI{XXXXXXX.XXXXXXX}

\acmPrice{15.00}
\acmISBN{978-1-4503-XXXX-X/18/06}

\usepackage{graphics}

\usepackage{enumitem}
\usepackage{makecell}
\usepackage{xspace}
\usepackage{nicematrix}
\usepackage{amsmath,amsfonts}

\usepackage{amsmath,amssymb,amsfonts}
\usepackage{graphicx}
\usepackage{textcomp}
\usepackage{subfigure}
\usepackage{tcolorbox}
\usepackage{booktabs}
\usepackage{tabularx}
\usepackage{lipsum}
\usepackage{multirow}

\newcommand{\later}[1]{}

\newcommand{\java}{Java\xspace}
\newcommand{\ruby}{Ruby\xspace}
\newcommand{\go}{Go\xspace}

\newcommand{\php}{PHP\xspace}
\newcommand{\python}{Python\xspace}
\newcommand{\javascript}{JavaScript\xspace}

\newcommand{\Our}{Telly-$K$\xspace} 
\newcommand{\OurBase}{Telly-} 
\newcommand{\layerfreezing}{layer\xspace freezing}

\newcommand{\codebert}{CodeBERT\xspace} 
\newcommand{\unixcoder}{UniXcoder\xspace} 
\newcommand{\codetf}{CodeT5\xspace}

\newcommand{\graphcodebert}{GraphCodeBERT\xspace}

\newcommand{\structural}{structural\xspace}
\newcommand{\Structural}{Structural\xspace}

\newcommand{\Fig}{Figure\xspace}

\newcommand{\Tab}{Table\xspace}
\newcommand{\Tabs}{Tables\xspace}
\newcommand{\Sec}{Section\xspace}
\newcommand{\Eq}{Eq.\xspace}

\usepackage{CJKutf8}
\usepackage[utf8]{inputenc} 

\usepackage{listings}
\usepackage{xcolor}
\definecolor{light-gray}{gray}{0.99}
\usepackage{fancybox}


\newcommand{\tabmargin}{-5pt}

\usepackage{arydshln}

\newcommand{\saveSpaceFig}{\vspace{-5pt}}
\newcommand{\savespacesec}{\vspace{0pt}}

\usepackage{tcolorbox}
\newcommand{\boxmargin}{0.15mm}
\tcbset{colback=gray!8,
        colframe=black,
        width=8.7cm,
        arc=2mm, auto outer arc,
        boxrule = 1.0pt,
        left = \boxmargin, right = \boxmargin, top = \boxmargin, bottom = \boxmargin,
        leftright skip=0.5mm
}

\begin{document}

\title[Towards Efficient Fine-tuning of Pre-trained Code Models]{Towards Efficient Fine-tuning of Pre-trained Code Models: \\ An Experimental Study and Beyond}

\newcommand\corrauthorfootnote[1]{%
  \begingroup
  \renewcommand\thefootnote{}\footnote{\textsuperscript{\S}#1}%
  \addtocounter{footnote}{-1}%
  \endgroup
}

\newcommand\notedauthorfootnote[1]{%
  \begingroup
  \renewcommand\thefootnote{}\footnote{\textsuperscript{\dag}#1}%
  \addtocounter{footnote}{-1}%
  \endgroup
}

\author{
Ensheng Shi\textsuperscript{a,\dag} \
Yanlin Wang\textsuperscript{b,\S,\dag} \
Hongyu Zhang\textsuperscript{c} \ \\
Lun Du\textsuperscript{d}\,
Shi Han\textsuperscript{d}\,
Dongmei Zhang\textsuperscript{d}\,
Hongbin Sun\textsuperscript{a,\S}
\\
\textsuperscript{a}Xi'an Jiaotong University \,
\textsuperscript{b}School of Software Engineering, Sun Yat-sen University \\
\textsuperscript{c}Chongqing University \quad
\textsuperscript{d}Microsoft  \quad
\\
{ s1530129650@stu.xjtu.edu.cn, wangylin36@mail.sysu.edu.cn, 
 hyzhang@cqu.edu.cn}\\
{  \{lun.du, shihan, dongmeiz\}@microsoft.com, hsun@mail.xjtu.edu.cn}\\
}
\renewcommand{\shortauthors}{Shi, et al.}

\settopmatter{printacmref=false}

\begin{abstract}
Recently, fine-tuning pre-trained code models such as CodeBERT on downstream tasks has achieved great success in many software testing and analysis tasks. While effective and prevalent, fine-tuning the pre-trained parameters incurs a large computational cost. In this paper, we conduct an extensive experimental study to explore what happens to layer-wise pre-trained representations and their encoded code knowledge during fine-tuning. We then propose efficient alternatives to fine-tune the large pre-trained code model based on the above findings. Our experimental study shows that (1) lexical, syntactic and structural properties of source code are encoded in the lower, intermediate, and higher layers, respectively, while the semantic property spans across the entire model. (2) The process of fine-tuning preserves most of the code properties. Specifically, the basic code properties captured by lower and intermediate layers are still preserved during fine-tuning. Furthermore, we find that only the representations of the top two layers change most during fine-tuning for various downstream tasks. (3) Based on the above findings, we propose \textbf{Telly} to efficiently fine-tune pre-trained code models via layer freezing. The extensive experimental results on five various downstream tasks demonstrate that training parameters and the corresponding time cost are greatly reduced, while performances are similar or better.
\end{abstract}

\vspace{-5pt}


%
\vspace{-10pt}

\maketitle
\savespacesec
\section{Introduction}

Recently, the pre-training with fine-tuning paradigm~\cite{AhmadCRC21,Lu2021,FengGTDFGS0LJZ20,GuoLDW0022} has achieved substantial improvement in many software testing and analysis tasks such as vulnerability detection~\cite{DingBPMRC22,GuoLDW0022}, patch generation~\cite{chakraborty2022natgen,GuoRLFT0ZDSFTDC21}, automatic program repair~\cite{JiangL021}, code review~\cite{prenner2021automatic,TufanoMMPPB22}, code generation~\cite{AhmadCRC21,chakraborty2022natgen,GuoLDW0022}, and clone detection~\cite{DingBPMRC22,GuoLDW0022,GuoRLFT0ZDSFTDC21}. They first pre-train large Transformer-base models to learn the general-purpose code representations on a large amount of data. Then, to adapt these models to the downstream tasks, they usually fine-tune them on targeted tasks~\cite{codet5wang,FengGTDFGS0LJZ20,GuoLDW0022,GuoRLFT0ZDSFTDC21,TufanoMMPPB22,NiuL0GH022,AhmadCRC21}.
 
In this paradigm, fine-tuning pre-trained code models usually achieves greatly better results on the downstream tasks. Although effective, fine-tuning the pre-trained parameters incurs a large computational cost with similarly large energy consumption. As reported in CodeXGLUE~\cite{Lu2021}, they usually require more than 10 hours to fine-tune the pre-trained model on a machine with two P100 cards for downstream tasks. 
In particular, as pre-trained models or fine-tuned datasets become larger, the computational cost becomes more expensive.
For example, \codetf~\cite{codet5wang} which has about 220MB of parameters spends over 40 hours fine-tuning it on CONCODE~\cite{iyer2018mapping} dataset for code generation. In fact, these actions are contrary to low-carbon deep learning~\cite{StrubellGM20}. 
In the software engineering area, there are only a few studies that explore what would happen to pre-trained code models during the fine-tuning process. Most related studies~\cite{FengGTDFGS0LJZ20,WanZZSXJ22,troshin2022probing,karmakar2021pre,hernandez2022ast} aim to understand what pre-trained code models know about source code. \textbf{There is a clear need to understand what happens to the pre-trained code models during fine-tuning and further efficiently adapt the pre-trained models to downstream tasks with less computational cost.}

In this paper, we\corrauthorfootnote{Yanlin Wang and Hongbin Sun are the corresponding authors.} first explore what code properties are encoded in layer-wise representations of pre-trained code models and what happens to these representations during fine-tuning\notedauthorfootnote{Work done during the author’s employment at Microsoft Research Asia.}. Then, we propose some efficient alternatives to fine-tuning for pre-trained code models based on the above findings. Specifically, first, inspired by the compilation process~\cite{aho2007compilers} and static program analysis techniques~\cite{moller2012static}, we propose four probing tasks (introduced in \Sec~\ref{sec:probing_task_intro}) involving the lexical, syntactical, semantic, and \structural code properties. Next, we conduct an empirical study to explore what code properties are encoded in pre-trained code models and what contributions of different layers are to the understanding of the encoded properties. Furthermore, we conduct an extensive experimental study to delve into what happens to layer-wise representations during fine-tuning on five diverse downstream tasks (shown in \Tab~\ref{tab:downstream-tasks}) including code search~\cite{GuZ018, shi2022enhancing}, clone detection~\cite{svajlenko2014towards,dang2011code}, code summarization~\cite{shia2022evaluation,wang2020cocogum}, code generation~\cite{IyerKCZ18}, and line-level code completion~\cite{Lu2021}. Through extensive experiments, we obtain the following major findings about pre-trained code models.

The \textit{first major finding} is that pre-trained code models encode the lexical property of source code mainly in the lower layers, recognize syntactical property mainly in the intermediate layers, and understand \structural property mainly in higher layers. The semantic properties are perceived across layers in the entire model. 
The \textit{second major finding} is that the process of fine-tuning preserves most of the code properties. 
That is, during fine-tuning, the basic code knowledge (or properties) encoded in lower and intermediate layers is still preserved. Only the knowledge captured by higher layers varies the most. In addition, our experimental study demonstrates that, when fine-tuning the pre-trained models on five diverse downstream tasks, the representations of lower layers change slightly, with only the top two layers showing substantial changes.

Based on the above findings, we propose \textbf{\Our}, for efficient fine-\underline{T}uning of  pr\underline{e}-trained code mode\underline{l}s via  \underline{l}a\underline{y}er freezing. Different $K$ values mean different variants of our approach. Specifically, we decrease the trained parameters via freezing the pre-trained parameters of the bottom $K$ layers that change insignificantly during fine-tuning, where $K \in [0,1,2,3, ..., L$-$1]$, the $0$-th layer is the embedding layer, and $L$ is the maximum number (typically 12) of layers of the pre-trained code model. Thus, \OurBase1 means freezing the embedding and the 1-st encoder layer. We conduct extensive experiments on five different downstream tasks from three aspects including training parameters, time cost, and performance. The evaluated pre-trained code models have 12 hidden layers. The experimental results show that (1) for almost all \Our $(0 \leq K \leq 11)$, the training time cost and parameters are substantially reduced, without significant changes in model performance.  
(2) When freezing the bottom $K (0 \leq K \leq 5)$ layers, training parameters are reduced by about 30\% to 65\%, and the training time is saved accordingly by about 10\% to 75\%. The model performance generally increases by 1\% to 4\% for different downstream tasks. (3) When freezing the bottom $K (6 \leq K \leq 9)$ layers, the training parameters are reduced by 65\% to 80\%, correspondingly saving about 50\% to 80\% of training time, while the model performance only changes slightly. (4) When the number of frozen layer is greater than nine $(10 \leq K \leq 11)$, training parameters and corresponding training time cost are tremendously reduced, while the model performance also drops significantly. 

Our main contributions are summarized as follows:
\begin{itemize}[topsep=0pt,itemsep=0pt,partopsep=0pt,parsep=0pt,leftmargin=10pt]
    \item  We propose four probing tasks related to lexical, syntactic, semantic, and \structural code properties. We explore what and how code properties are encoded in layer-wise representations through the above probing tasks.  

     \item To the best of our knowledge, we are the first to conduct an extensive experimental study to analyze what happens to layer-wise representations and their encoded code properties during fine-tuning of pre-trained code models. 

     \item We propose an efficient approach to fine-tune pre-trained code models to downstream tasks via \layerfreezing. In addition, we conduct extensive experiments on five different downstream tasks to demonstrate the efficiency of our approach. 
\end{itemize}

The rest of this paper is organized as follows.~\Sec~\ref{sec:background} introduces the relevant background knowledge. Then, we conduct an experimental study to understand what happens to pre-trained code models during fine-tuning in ~\Sec~\ref{sec:experimental_study}. Based on the above findings, in~\Sec{}~\ref{sec:efficient_fine_tuning}, we propose \Our{} and conduct the extensive experiments on five different downstream tasks to show its superiority.~\Sec~\ref{sec:discuss_threats} discusses the importance of reducing fine-tuning time, actionable guideline to better fine-tuning, and the generality of our experimental findings, and identifies some threats to validity. ~\Sec{}~\ref{sec:related_work} presents related work. Finally, we summarize our paper and discuss the future work in \Sec~\ref{sec:conclusion}.

\savespacesec
\section{Background}
\label{sec:background}
\subsection{Pre-trained Code Models}
Large pre-trained models have achieved substantial results in many areas including natural language processing~\cite{BertDevlinCLT19,Liu2019Roberta}, computer vision~\cite{MAEHeCXLDG22,Bao0PW22} and software engineering~\cite{AhmadCRC21,FengGTDFGS0LJZ20,GuoLDW0022,GuoRLFT0ZDSFTDC21,ZengTZLZZ22}. In the software engineering community, generally, they firstly pre-train large models on amounts of 
source code-related data, and then fine-tune them on downstream tasks to improve their performance. Recently, many pre-trained code models~\cite{FengGTDFGS0LJZ20,GuoLDW0022,GuoRLFT0ZDSFTDC21,codet5wang,NiuL0GH022,AhmadCRC21} have been proposed and shown the surprisingly promising results on many software engineering tasks involving software testing,  security, maintenance, and development~\cite{Lu2021,DingBPMRC22,GuoLDW0022,prenner2021automatic,TufanoMMPPB22}. We introduce these models from three aspects as follows.

\textbf{Basic architecture.} Most recent pre-trained code models adopt the multi-layer Transformer model~\cite{Transformer17} (typically, a Transformer encoder) as the basic architecture. A Transformer encoder is essentially composed of an embedding layer, a positional encoder, and a stack of encoder layers. In general, given an input code snippet, it is firstly 
embedded by the embedding layer and the positional encoder to obtain the initial word embeddings. Next, they are fed to the multiple stacked encoder layers to encode the input information layer by layer. 

Mathematically, we denote the input tokens as $T = [t_1, t_2, ..., t_n]$, where $n$ is the length of the input token sequence. The embedding layer maps each token to a high-dimensional semantic space.  The positional encoder is used to encoder the positional information and then injected it into input embedding by:
\begin{equation}
\small
    w_i = embed(t_i) + pos(t_i), \quad i = 1, 2, ... ,n
\label{eq:input_embedding}
\end{equation}
\noindent where $embedd(*)$ and $pos(*)$ denote the embedding layer and positional encoder, respectively.
$W = [w_1, w_2, ... , w_n] $ is the initial word embeddings. Next, the multiple stacked encoder layers produce a set of layer-wise contextual representations $H^{0}, H^{1}, ..., H^{L}$ by:
\begin{equation}
\begin{aligned}
\small
    H^{0} &= [w_1, w_2, ... , w_n] \\
    H^{l} &= encoder^{l}(H^{l-1}), \quad l = 1,2, ..., L    
\end{aligned}
\end{equation}
where $L$ is the number of stacked layers, and $encoder^{l}(*)$ denotes the $l$-th encoder layer. $H^{l}= [h^{l}_{1}, h^{l}_{2}, ..., h^{l}_{n}]$ denotes the contextual representations of the  $l$-th layer, and $H^{0}$ is the initial word embeddings.

\textbf{Pre-training.} 
Pre-training techniques, as one of the self-supervised learning approaches, can leverage a big model to learn the general representations with amounts of unlabeled dataset~\cite{BertDevlinCLT19,Liu2019Roberta,RaffelSRLNMZLL20,MAEHeCXLDG22}. Typically, such techniques usually automatically generate virtual labels from the unlabeled samples to reformulate an unsupervised learning problem as one that is supervised learning. The corresponding supervised tasks are named pre-trained tasks. 
For example, \codebert~\cite{FengGTDFGS0LJZ20} uses a 12-layer Transformer encoder with 768-dimensional embedding and pre-trains all parameters on a large-scale dataset named CodeSearchNet~\cite{CSNHusain2019} (contains 2.1M bimodal data (code functions paired with natural language comments) and 6.4M unimodal codes across six programming languages (\ruby{}, \javascript{}, \go, \python, \java, \php) with two pre-trained tasks, namely, masked language modeling and replaced token detection. The former task is to predict the original tokens of the masked positions, while the latter is to identify whether a token is the original one or not.
\unixcoder~\cite{GuoLDW0022} takes code/text sequence as input and is pre-trained on the C4 dataset from T5~\cite{RaffelSRLNMZLL20} and 4.1M unimodal code from CodeSearchNet with five different pre-trained tasks.

\textbf{Fine-tuning.} After pre-training on the massive dataset, they adapt pre-trained models to downstream tasks by fine-tuning all the pre-trained parameters on the targeted dataset~\cite{FengGTDFGS0LJZ20,GuoLDW0022,GuoRLFT0ZDSFTDC21,codet5wang,NiuL0GH022,AhmadCRC21,du2021single}. For example, in code search, previous studies~\cite{GuoLDW0022,Lu2021} average the last-layer contextual representations (e.g. $H^{L}$) of models as the overall representation,  measure the similarity between representations of the source code and the query by vector distance, and fine-tune them by pulling together the paired code and query and pushing apart the unpaired code and query. Compared with pre-trained models, the MRR values of fine-tuned models on code search are improved from 0.001 and 0.156 to 0.694 and 0.713 for \codebert and \graphcodebert, respectively.
\savespacesec
\vspace{-10pt}
\subsection{Probing Techniques}
Probing techniques have been extensively used in the NLP community to study what linguistic properties are captured by pre-trained language models. Specifically, they extract contextual representations (such as  $H^{1}$,$H^{2}$ ) from the pre-trained model as frozen features, feed them to a linear classifier, and only train it to predict probing tasks relevant to linguistic properties. They also take random representations as a baseline to demonstrate the ability of the pre-trained representations to encode linguistic attributes for comparison. For example, Tenney et al.~\cite{TenneyXCWPMKDBD19} utilize different probing tasks such as part-of-speech tagging, dependency parsing, semantic role labeling, and coreference resolution labeling to examine the ability of pre-trained models to understand linguistic properties, such as parts of speech, dependencies, semantics, and coreferences. In software engineering area,  most related studies~\cite{WanZZSXJ22,karmakar2021pre,hernandez2022ast} aim to understand what pre-trained code models know about source code. For example, Wan et al.~\cite{WanZZSXJ22} propose a new probing task to investigate whether the structure of ASTs is encoded in the representations of pre-trained code models. Specifically, they extract layer-wised pre-trained representations of two input code tokens from code pre-trained models, feed them to a matrix, and train the matrix to reconstruct the distance between two corresponding terminal nodes in the AST, where the distance is used to represent the syntactical structure.
López et al~\cite{hernandez2022ast} propose a new probing task named AST-Probe, which recovers ASTs from hidden representations of pre-trained language models. Specifically, AST-probe first maps the layer-wise representations of pre-trained code models to a latent space, called syntactic subspace, using the orthogonal projection and then uses geometry of this space to predict the AST of the input code snippet. Karmakar et al~\cite{karmakar2021pre} also propose some new probing tasks to investigate whether pre-trained language models can understand some simple code properties. 
Motivated by the compilation process~\cite{aho2007compilers} and static analysis techniques~\cite{moller2012static}, we propose four probing tasks (\Sec~\ref{sec:probing_task_intro}) related to different code properties.
Besides probing pre-trained code models, we also study their layer-wise representations during fine-tuning.

\savespacesec
 \subsection{Representational Similarity Analysis}
\label{sec:background-rsa}
Representational similarity analysis (RSA) is originally used in cognitive neuroscience~\cite{kriegeskorte2008representational} to study the relation between the neural activation patterns in human brains and representations of a computational model for given a set of stimuli. Recently, it is adopted to measure the similarity between two representational spaces~\cite{AbnarBCZ19, ChrupalaA19, MerchantRPT20}. For example, given a set of inputs, different models generating different representational spaces would generate different representations. Merchant et al.~\cite{MerchantRPT20} construct two distance matrices. Each records the vector distances (such as cosine similarity) between representations in one representational space. Then the representational similarity of two representational spaces is measured by the Pearson’s correlation~\cite{schober2018correlation} of these two distance matrices. In general, a value of the correlation coefficient between 0.8 and 1 indicates that the two representational spaces are fairly similar, while a value lower than 0.5 means the two representational spaces are dissimilar~\cite{akoglu2018user,schober2018correlation}. In this paper, we conduct representational similarity analysis to study the similarity of layer-wise representations between pre-trained and fine-tuned models in \Sec~\ref{sec:rsa_pipline} and~\ref{sec:findings_rq1_rq2}

\savespacesec
\section{An Experimental Study on Pre-trained Code Model}
\label{sec:experimental_study}

\subsection{Research Questions}
While effective and prevalent, fine-tuning pre-trained code models incur a  large computational cost. In this work, we first conduct an experimental study to investigate what code properties are encoded in layer-wise pre-trained representations and what happens to these representations during fine-tuning.  
The research questions on the experimental study are introduced in detail as follows.

\paragraph{RQ1: What code properties are encoded in layer-wise pre-trained representations?}
Probing techniques have been extensively used in the NLP community to analyze and interpret pre-trained language models. Motivated by the compilation process~\cite{aho2007compilers} and static analysis techniques~\cite{moller2012static}, we first propose four probing tasks related to lexical, syntactic, semantic, and \structural properties of source code. They are introduced in detail in \Sec~\ref{sec:probing_task_intro}. Then, we investigate what code properties are encoded by layer-wise pre-trained representations via the above probing tasks in \Sec~\ref{sec:probing_pipeline}. At the same time, we study how much representations of each layer contribute to understanding these code properties. Further, we perform the same probing experiments for the model fine-tuned in a downstream task in \Sec~\ref{sec:findings_rq1_rq2}, and intuitively understand what happens to the code properties captured by the pre-trained model during fine-tuning.

\paragraph{RQ2: What happens to the layer-wise representations during fine-tuning?}
In the RQ1, we aim to roughly understand what happens to a pre-trained code model when fine-tuning with the help of probing tasks. We further conduct the extensive representation similarity analysis (RSA) to study what happens to pre-trained representations layer-by-layer when fine-tuning them on downstream tasks.
RSA introduced in \Sec~\ref{sec:background-rsa} is a task-agnostic technique and requires no prior knowledge of probing tasks. How to apply RSA to the pre-trained and fine-tuned models is described in \Sec~\ref{sec:rsa_pipline}. To ensure the generality of our experimental findings, we conduct experiments on five diverse downstream tasks including code search, clone detection, code summarization, code generation, and line-level code completion. 

\subsection{Probing Pre-trained Code Models}
\begin{figure*}[t]
     \centering
    \subfigure[Source code snippet]{
    \begin{minipage}{0.26\linewidth}
    \centering
    \includegraphics[width=0.6\linewidth]{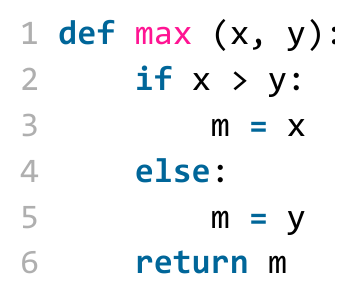}
    \vspace{4pt}
    \label{fig:probing:code}
    \end{minipage} 
    }
    \subfigure[Syntactic Probing]{
    \begin{minipage}{0.4\linewidth}
        \centering
         \vspace{3pt}
        \includegraphics[width=1.0\linewidth]{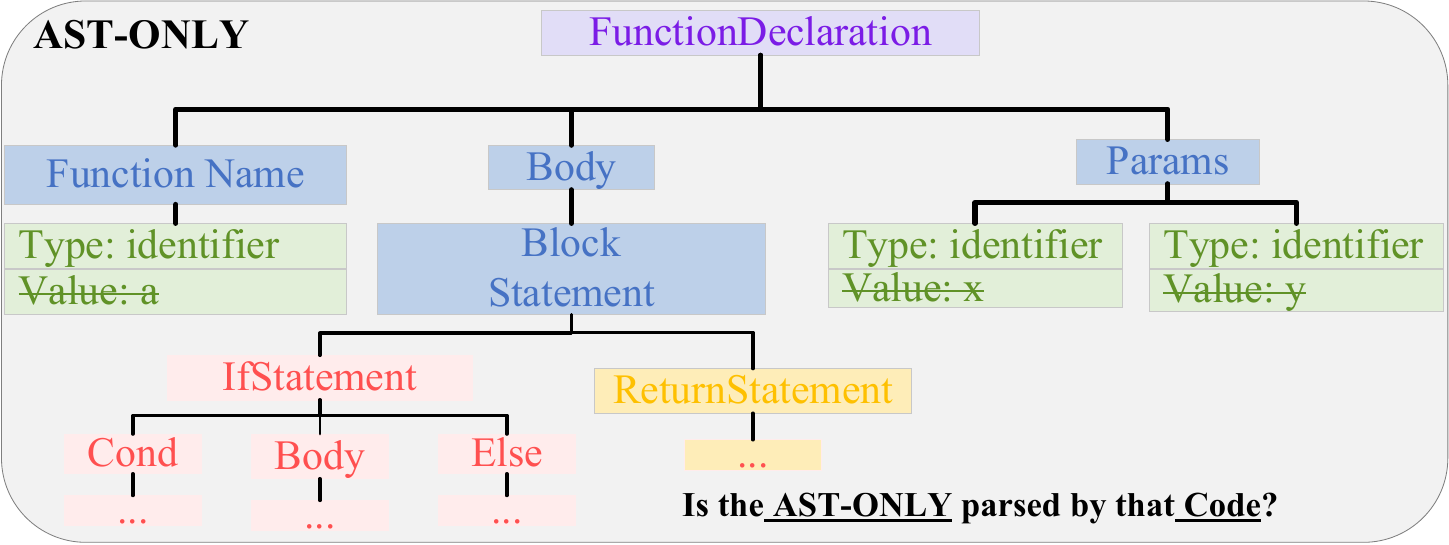}
      \vspace{-2pt}
    \end{minipage}
    \label{fig:probing:syntactic}
    }
            \subfigure[\Structural probing]{
    \begin{minipage}{0.3\linewidth}
         \centering
        \includegraphics[width=0.6\textwidth]{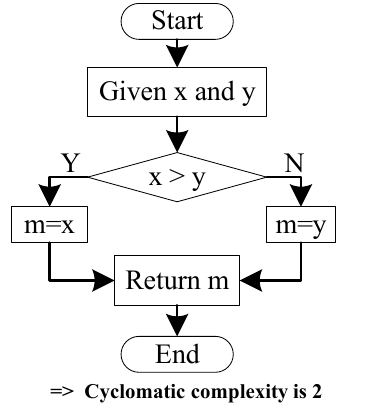}
          \vspace{1pt}
    \end{minipage}
   
    \label{fig:probing:structural}
    }

    \vspace{-15pt}
    
    \subfigure[Lexical probing]{
    \begin{minipage}{0.45\linewidth}
        \centering
          \includegraphics[width=0.94\linewidth]{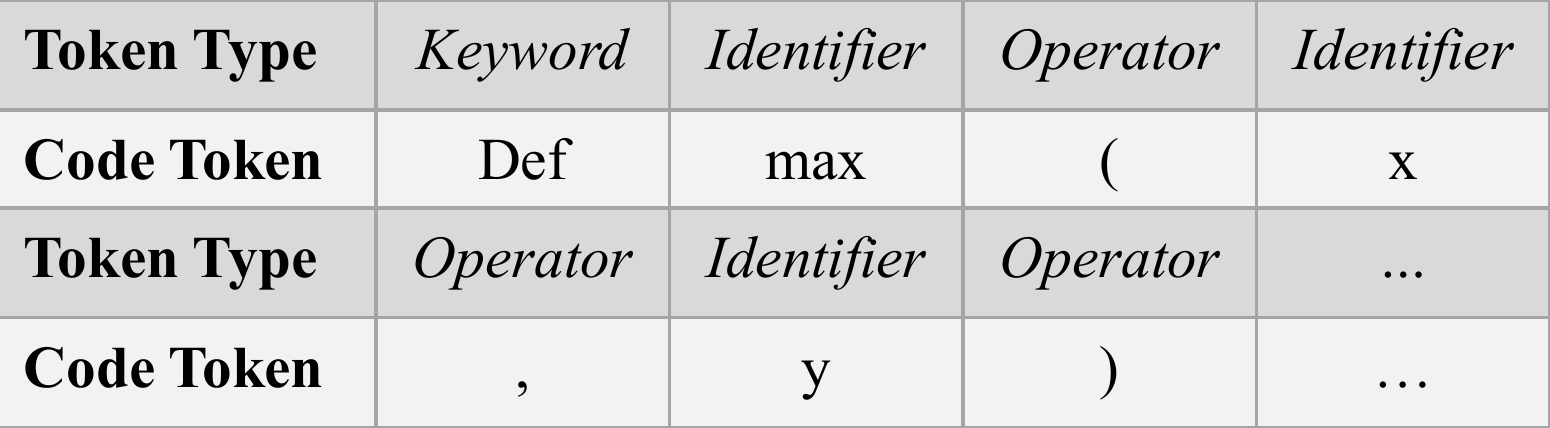}
             \vspace{1pt}
    \end{minipage}
 
    \label{fig:probing:lexical}
    }
    \subfigure[Semantic probing]{
    \begin{minipage}{0.45\linewidth}
        \centering
        \includegraphics[width=0.99\textwidth]{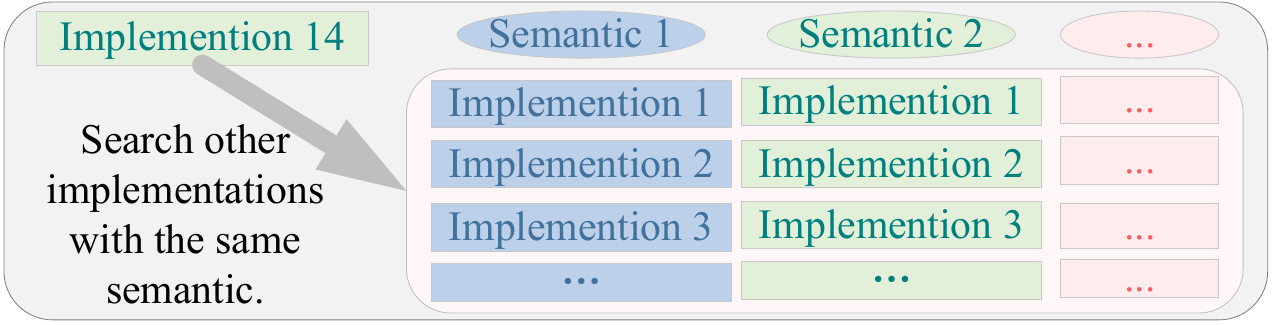}
        \vspace{1pt}
    \end{minipage}

    \label{fig:probing:semantic}
    }
\saveSpaceFig
\caption{An example of source code and probing tasks}
\label{fig:probing_task}
\end{figure*}

We introduce the four code-related probing tasks and probing pipeline as follows.

\savespacesec
\subsubsection{Four probing tasks}
\label{sec:probing_task_intro}
We design and show the four probing tasks in \Fig~\ref{fig:probing_task}. They are related to lexical, syntactical, semantic and \structural code properties. We introduce them one by one in detail.

\textbf{Lexical probing.} Lexical probing aims to measure how well contextual representations encode the lexical properties of source code. As we all know, when the source code is compiled, the first step is lexical analysis, which tokenizes the source code string and determines the type (such as \textit{Identifier, Keywords}) of each code token. Different types play very different semantic or syntactic roles in subsequent program analysis and compilation. Thus, it is important to understand whether pre-trained code models have captured the lexical information of source code by contextual representations. To achieve this, we first use the contextual representations of pre-trained code models as frozen features, then feed them to a linear classifier, and finally train it to predict the type of each code token. As shown in \Fig~\ref{fig:probing:lexical}, each token belongs to one of the five types including \textit{Identifier, Keyword, Operator, Number, and String}. Due to space limitation, the detailed definition of each type and description of lexical probing can be found in the online Appendix of the replication package~\cite{ISSTA2023Repo}.
   
\textbf{Syntactic probing.} Syntax analysis typically comes after the lexical analysis in the process of program compilation~\cite{aho2007compilers}, where a parser takes token sequence generated by the lexer as input and produces data structures like parse tree or abstract syntax tree (AST). Similarly, syntactic probing is designed to investigate how well contextual representations perceive the syntactic properties of source code. The basic idea is to identify whether a code and an anonymous AST (named AST-Only shown in \Fig~\ref{fig:probing:syntactic}) are paired. Specifically, we first parse the source code to obtain the corresponding AST, which includes the non-terminal and terminal nodes. The non-terminal nodes represent the syntactic information, while terminal nodes consist of types corresponding to the syntactic elements and values corresponding to code tokens in the source code. It is easy for one model to identify whether the AST is parsed by one code snippet according to the overlap of code tokens. Therefore, as shown in \Fig~\ref{fig:probing:syntactic}, we construct the AST-Only by removing the values of terminal nodes. Next, we train a linear classifier to determine whether the given AST-Only is parsed from the given code snippet or not. 
The true pairs are constructed by pairing the code with the corresponding parsed AST-Only, while false pairs are constructed by pairing the code with an AST-Only parsed by other different codes. Detailed descriptions of syntactic probing can be found in the online Appendix~\cite{ISSTA2023Repo}.

 \begin{table*}[ht]

\caption{An overview of downstream tasks, which includes descriptions, and evaluated datasets, programming languages and metrics. \#Size shows the sizes of train, validation and test sets in order. For metrics, P, R, F1 are short for precision, recall and F1-score, respectively. EM and Edit sim are short for Exact Match accuracy and Levenshtein edit similarity, respectively.}
\label{tab:downstream-tasks}
 \vspace{-10pt}
\small
  \setlength{\tabcolsep}{4.5pt}
\renewcommand{\arraystretch}{0.85}
\begin{tabular}{llllll}
\toprule 
Task &\multicolumn{1}{c}{Description}  & Dataset Name & Language & \multicolumn{1}{c}{\#Size} &\multicolumn{1}{c}{Metrics} \\
\midrule 
 \multirow{2}{*}{Code search} & \multirow{2}{*}{\shortstack[l]{\\Search semantically relevant code snippets \\ for a given natural language query.}} & \multirow{2}{*}{CodeSearchNet~\cite{CSNHusain2019} }&\python & 251K/9.6K/1K & MRR, R@1  \\
 \cmidrule(r){4-5}
&& & \ruby & 24.9K/1.4K/1.3K  &  R@5, R@10\\
\midrule
\multirow{2}{*}{Clone detection}&\multirow{2}{*}{\shortstack[l]{Detect whether two code snippets \\ are functional equivalence.}} &\multirow{2}{*}{BigCloneBench~\cite{svajlenko2014towards}} & \multirow{2}{*}{Java} & \multirow{2}{*}{901K/416K/416K} &\multirow{2}{*}{P, R, F1}\\
\\
\midrule
 \multirow{2}{*}{Code summarization}& \multirow{2}{*}{\shortstack[l]{\\Generate the concise natural language \\ description for the given code snippet.}} & \multirow{2}{*}{CodeSearchNet~\cite{CSNHusain2019} }&\python & 251K/9.6K/1K &BLEU, Meteor, \\
 \cmidrule(r){4-5}
&& & \ruby &24.9K/1.4K/1.3K &Rouge-L,Cider \\
\midrule
\multirow{2}{*}{Code generation}&\multirow{2}{*}{\shortstack[l]{Generate  a function-level code snippet for\\ the given natural language description.}} &\multirow{2}{*}{CONCODE~\cite{iyer2018mapping}} & \multirow{2}{*}{Java} & \multirow{2}{*}{100K/2K/2K }&\multirow{2}{*}{BLEU, EM} \\
\\
\midrule
\multirow{2}{*}{Code completion}&\multirow{2}{*}{\shortstack[l]{ Predict the next line of code for the given \\ previous code context.}} &\multirow{2}{*}{ Github Java Corpus~\cite{allamanis2013mining}} & \multirow{2}{*}{Java} & \multirow{2}{*}{12K/1.5K/1.5K} &\multirow{2}{*}{Edit sim, EM}\\
\\
\bottomrule
\vspace{-10pt}
\end{tabular}
\end{table*} 

\textbf{Semantic probing.} To understand to what extent pre-trained code models are aware of code semantics, we perform semantic probing (\Fig~\ref{fig:probing:semantic}), which examines the ability to identify code snippets with the same semantics but different implementations. Specifically, we use POJ-104~\cite{mou2016convolutional} dataset, which consists of 104 problems and 500 C/C++ implementations for each problem, as the evaluated dataset. We train a linear mapper taking the pre-trained representations as input to map semantically similar code snippets into similar embeddings. Thus, the implementations with the same semantic can be easily recalled by the vector distance of them. Detailed descriptions of semantic probing can be found in the online Appendix~\cite{ISSTA2023Repo}.
 
\textbf{\Structural probing.} In addition to lexical, syntactic, and semantic properties, \structural properties are also important for code analysis. 
Cyclomatic complexity~\cite{mccabe1976complexity}, which indicates the complexity of a program and can be referred to control flow graph (CFG), can be used as a \structural property of code. Mathematically, the cyclomatic complexity $M$ can be  calculated based on the CFG of source code by:
\begin{equation}
\small
    M=E-N+2P
\end{equation}
where $E$ and $N$ are the number of edges and nodes of the graph, respectively. $P$ is the number of connected components. Its value is typically 1 because the CFG is a connected graph. As shown in \Fig~\ref{fig:probing:structural}, the CFG has 7 nodes and 7 edges, hence the cyclomatic complexity of the code snippet is $7-7+2 =2$. We use the cyclomatic complexity prediction as the \structural probing task to investigate how well contextual representations understand the \structural property of source code. Detailed descriptions of \structural probing can be found in the online Appendix~\cite{ISSTA2023Repo}.

\savespacesec
\subsubsection{Probing pipeline}
\label{sec:probing_pipeline}
Following previous studies~\cite{TenneyDP19,MerchantRPT20}, to investigate what code properties are encoded in layer-wise representations of the pre-trained model, we train a classifier that takes these layer-wise representations as input to predict the probing task associated with one of code properties.
At the same time, we learn a linear combination of contextual representations of all layers to
study how much representations of each layer contribute to understanding these code properties. Mathematically, for the pre-trained layer-wise representations $H^{0},H^{1},...,H^{L}$, we combine them by:
 \begin{equation}
\begin{aligned}
\small
F &=\sum_{l=1}^{L} \lambda^{l} H^{l}, \quad
\lambda^{l} = \frac{\exp a_{l}}{\sum_{i=0}^{L} \exp a_{i}} \\
\end{aligned}
\label{eq:layer_weights}
\end{equation}
where the layer-wise weight $a_{l}$ are jointly learned with the probing classifier. On the one hand, we compare the performance between combined representations $F$ and randomly initialized representations to study how well the pre-trained contextual representations encode the properties of source code. We also compare the performance between pre-trained and fine-tuned representations to study what happens to the code properties during fine-tuned. On the other hand, to investigate how much the representation of each layer contributes to encoding a code property and to explore the differences between pre-trained and fine-tuned models, we present layer-wise weight $a_{l}$ of pre-trained and fine-tuned models for each probing task and further analyze the experimental results ~\Sec~\ref{sec:findings_probing}.

\savespacesec
\subsection{Representational Similarity Analysis}
\label{sec:rsa_pipline}
Following the previous study~\cite{MerchantRPT20}, we randomly sample $N$ code snippets and obtain the layer-wise representations of the pre-trained and fine-tuned model. Then, for each layer, we obtain distance matrix $A^{l}$ (introduced in Section~\ref{sec:background-rsa} and the size is $ N \times N)$ for the $l$-th layer by calculating the cosine similarity between this layer's representational vectors of 
any two code snippets. Representational vectors are obtained by averaging that layer's contextual representations. Mathematically,  we denote the $l$-th contextual representations of the pre-trained or fine-tuned code model for the $k$-th code snippet as ${H_{k}}^l$.
The distance matrix $A^l$ is calculated by:
\begin{equation}
\small
     \quad A^l_{i,j} = \frac{v^l_i \cdot v^l_j }{\Vert v^l_i\Vert\Vert v^l_j\Vert}, \quad v^l_k = mean({H_{k}}^l),\quad i,j,k \in [1, 2, ..., N] 
\end{equation}
Next, for the $l$-th layer, we calculate the Pearson’s correlation coefficients $\rho^{l}$ between the two distance matrices obtained from pre-trained and fine-tuned models, respectively. In particular, we conduct experiments on five diverse downstream tasks including code search, clone detection, code summarization, code generation, and line-level code completion. The overview of them is in \Tab~\ref{tab:downstream-tasks}.  The experimental results are shown in~\Sec~\ref{sec:rq2_result}.

\savespacesec

\subsection{Experimental Settings}
In this study, we analyze the state-of-art pre-trained code models~\unixcoder~\cite{GuoLDW0022} and ~\graphcodebert~\cite{GuoRLFT0ZDSFTDC21}. Both of them are 12-layer Transformer with 768 dimensions and the total parameters are about 120 MB. \unixcoder is a unified pre-trained code model and can be used as an encoder, a decoder, or an encoder-decoder architecture by a special indication token. \graphcodebert considers the data flow information 
and pre-trains a large model using a lot of bimodal data (code functions paired with natural language comments) and unimodal codes data. We conduct the experiment on \unixcoder and \graphcodebert because they all achieve promising results on many code intelligence tasks.

For the experiments on probing, we construct the evaluation datasets through CodeSeachNet and POJ-104~\cite{mou2016convolutional} shown in \Tab~\ref{tab:downstream-tasks}. The lexical, syntactic, and \structural probings employ the CodeSeachNet dataset with \python, and semantic probing uses the POJ-104~\cite{mou2016convolutional}.
Following the fine-tuning experimental settings of \unixcoder /\graphcodebert on code search, We fine-tune the pre-trained model on CodeSeachNet dataset with \python, and probe the pre-trained and fine-tuned model with the four probing tasks. When probing, the maximum length of code snippets is set to 512.  The maximum epoch and batch size are set to 30 and 32, respectively. We adopt the Adam optimizer with a learning rate of 1e-4 and perform early stopping on the validation set. We run the experiments 3 times with random seeds 0,1,2 and display the mean value in the paper.

For the representational similarity analysis, following the previous study~\cite{MerchantRPT20}, $N$ is set to 5,000. Following the fine-tuning experimental settings of \unixcoder /\graphcodebert, we fine-tune it on the five downstream tasks shown in \Tab~\ref{tab:downstream-tasks}. 

\savespacesec
\vspace{-3pt}
\subsection{Experimental Findings}
\label{sec:findings_rq1_rq2}
In this section, we present and analyze the results of the above two research questions. We present the results of \unixcoder-based \Our only due to space limitation and put results of \graphcodebert-based \Our in the online Appendix~\cite{ISSTA2023Repo}. Conclusions and findings that hold on \unixcoder generally hold for \graphcodebert.

\savespacesec
\subsubsection{RQ1: What code properties are encoded in layer-wise pre-trained representations?}
\label{sec:findings_probing}

\begin{table}[t]
 \setlength{\tabcolsep}{8pt}
    \centering
    \small
    \caption{The performance of probing tasks for random, pre-trained and fine-tuned representations.}
    \vspace{-10pt}
    \label{tab:probing}
    \begin{tabular}{l c c c }
    \toprule
    \multirow{2}{*}{Probing Task}  & \multicolumn{3}{c}{Performance} \\
      \cmidrule(r){2-4}
    & Random  & Pre-trained & Fine-tuned \\
    \midrule
     Lexical probing    &76.14 &99.98 &99.95\\
     Syntactic probing  &64.70  &95.60 &95.10 \\
     Semantic probing  &46.10 &73.25 &71.55\\
     Structural probing  &34.80 &89.80 &62.20\\
    \bottomrule
    \end{tabular}
    \vspace{-10pt}
\end{table}

\begin{figure}[t]
    \centering
    \includegraphics[width=0.9\linewidth]{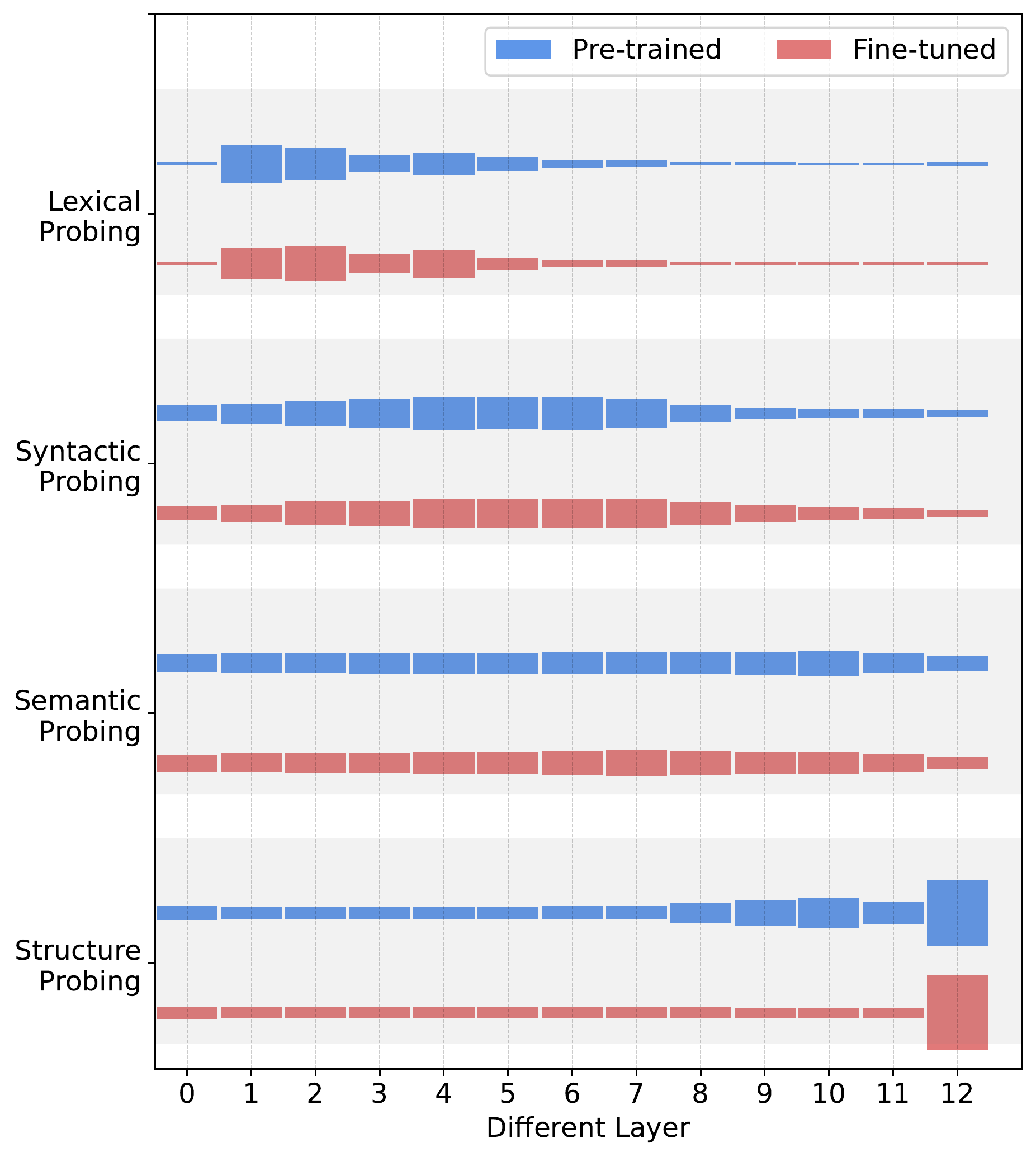}
     \vspace{-10pt}
    \caption{Layer-wise contributions on different probing tasks for the pre-trained and fine-tuned code model.}
    \label{fig:probing_weight_pre_train_fune_tuning}
     \vspace{-10pt}
\end{figure}

We use the four probing tasks related to lexical, syntactic, semantic and \structural properties to explore what code properties are encoded in layer-wise pre-trained representations and how much representations of each layer contribute to understanding these code properties. At the same time, we also compare pre-trained and fine-tuned layer-wise representations in the same setting.
The performance on probing tasks is shown in \Tab~\ref{tab:probing} and the layer-wise contribution are presented in \Fig~\ref{fig:probing_weight_pre_train_fune_tuning}.

In~\Tab~\ref{tab:probing}, the results of lexical, syntactic, and structural probing are measured by accuracy. The results of semantic probing are measured by mean average precision (MAP)~\cite{sanderson2010christopher}. 
For more detailed descriptions of the accuracy and MAP metrics, please refer to the online Appendix~\cite{ISSTA2023Repo}. From \Tab~\ref{tab:probing}, we can find that (1) pre-trained and fine-tuned representations better understand code properties than random representations; (2) after fine-tuning, lexical, syntactic and semantic code properties are still well captured, while the ability to capture the \structural property significantly declines.
The first finding is expected because pre-trained code models can take advantage of larger datasets and model sizes to encode the basic code knowledge into their representations. After fine-tuning, some code properties are still preserved. This may be related to the characteristics of downstream tasks since code search mainly relies on the lexical, syntactic, and semantic information of the code rather than the structural information. It may also be the result of vanishing gradients~\cite{bengio1994learning} because code properties encoded in the lower layer change very little, and code properties encoded in the higher layer change obviously. Actually,  vanishing gradients has little effect on the optimization of the pre-trained code model as the basic architecture employed by the model uses the residual connection~\cite{resnet16,Transformer17} which can effectively avoid the vanishing problem.

\Fig~\ref{fig:probing_weight_pre_train_fune_tuning} displays the layer-wise contributions ($\lambda^{l}$  in \Eq~\ref{eq:layer_weights}) for the pre-trained and fine-tuned model. From \Fig~\ref{fig:probing_weight_pre_train_fune_tuning}, 
\textit{on the one hand}, we can observe that for pre-trained or fine-tuned models, lexical, syntactic, and \structural properties of source code are mostly captured in the lower, intermediate, and higher layers, respectively, while the semantic property almost spans across the entire model. We conduct the significance testing\footnote{The results are put in the online Appendix~\cite{ISSTA2023Repo} due to space limitation} to examine the significance of the contribution differences. The experimental result shows that for lexical probing, the contributions of the 1-th, 2-th, and 4-th layers are significantly greater than other layers. For syntactic probing, the contributions of the 4-th to 7-th layers are significantly greater than other layers. For semantic probing, the contributions among different layers are not significantly different. For \structural probing, the contribution of the last layers of the pre-trained and fine-tuned model is significantly greater than others.  
\textit{On the other hand}, we find that the process of fine-tuning preserves most of the code properties. Specifically, basic code properties captured by lower and intermediate layers are still preserved during fine-tuning. Only the performance of \structural probing task changes obviously. 
 \begin{tcolorbox}
\textbf{Summary.} For pre-trained layer-wise representations, lexical, syntactic, and \structural properties of source code are mainly captured by 
the lower, intermediate and, higher layers, respectively, while the semantic property almost spans across the entire model. Meanwhile, the basic code properties captured by lower and intermediate layers are still preserved during fine-tuning. 
\end{tcolorbox}

\subsubsection{RQ2: What happens to the layer-wise representations during fine-tuning?}
\label{sec:rq2_result}
\begin{figure}[t]
    \centering
    \includegraphics[width=1.0\linewidth]{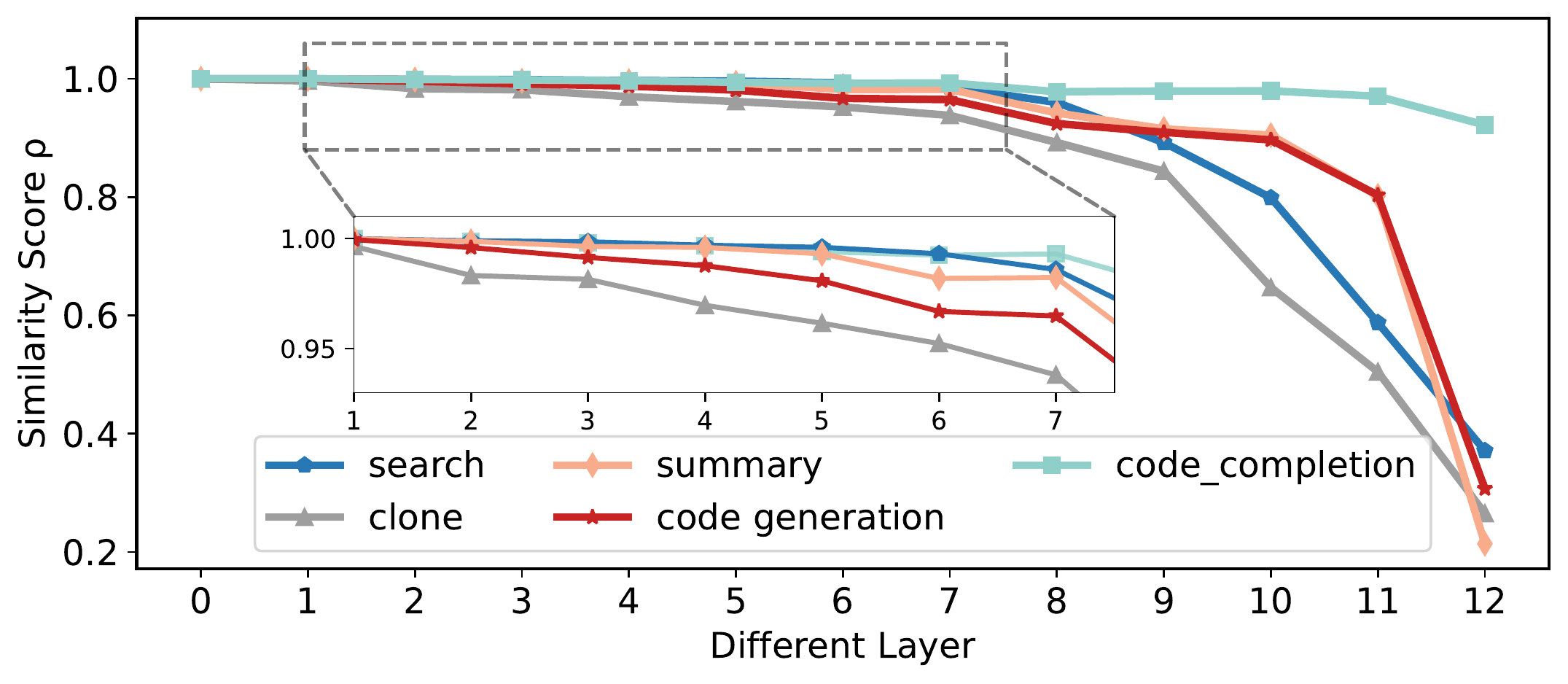}
    \vspace{-20pt}
    \caption{Similarity scores between pre-trained and fine-tuned models for five downstream tasks. }
    \label{fig:rsa_search_py}
    \vspace{-10pt}
\end{figure}
We also conduct extensive experiments on representational similarity analysis (RSA) to study what happens to the layer-wise representations of the pre-trained model during the fine-tuning for five diverse downstream tasks without the help of probing tasks. The results are shown in \Fig~\ref{fig:rsa_search_py}.
From the presented results, we can see that the correlation coefficients (similarity scores in \Fig~\ref{fig:rsa_search_py}) of the bottom 9 layers are all greater than 0.8. It means that representations of the bottom 9 layers are similar between pre-trained and fine-tuned models for the five downstream tasks. The representations of the top layer are dissimilar  ($\rho \leq 0.5$) except for code completion. This is because the top layer of the pre-trained model \unixcoder are used to predict the masked tokens, which is similar to the experimental setting of code completion. Furthermore, we find that the representation of the bottom 7 layers ($\rho \geq 0.9$) are greatly related and the bottom 5 layers ($\rho \geq 0.95$) are strongly similar.

\begin{tcolorbox}
\textbf{Summary.} The representations of the bottom nine layers are similar between the pre-trained and fine-tuned models for the five downstream tasks. Only the representations of the top two layers change greatly during fine-tuning.
\end{tcolorbox}

\vspace{-10pt}
\section{Efficient Fine-tuning of Pre-trained Code Models}
\label{sec:efficient_fine_tuning}
\begin{table*}[t]
    \centering
    \small
    \caption{Experimental results on code search. \#Params is short for the number of training parameters. M is short for million. The changing ratios compared to the base model are shown in parentheses.}
    \vspace{-8pt}
    \label{tab:freeze_on_search}
    \begin{tabular}{cccccccccc}
    \toprule
    \multirow{2}{*}{Model} & \multirow{2}{*}{Note} & \multirow{2}{*}{\#Params } & \multicolumn{2}{c}{Training time} & \multicolumn{4}{c}{Performance} \\
    \cmidrule(r){4-5} \cmidrule(r){6-9} 
    & & &Each epoch &Convergence &MRR &R@1 &R@5 &R@10 \\ 
    \midrule
Base &Fine-tuning all parameters &125.93M&17m14s&2h35m06s&0.720&0.612&0.838&0.889\\
\midrule
\OurBase0 &Freezing the bottom 0 layers &85.0M($\downarrow$32\%)&15m49s($\downarrow$8\%)&2h06m32s($\downarrow$18\%)&0.727($\uparrow$1\%)&0.630($\uparrow$3\%)&0.846($\uparrow$1\%)&0.895($\uparrow$1\%)\\
\OurBase1 &Freezing the bottom 1 layers &78.0M($\downarrow$38\%)&14m53s($\downarrow$14\%)&1h59m04s($\downarrow$23\%)&0.727($\uparrow$1\%)&0.629($\uparrow$3\%)&0.848($\uparrow$1\%)&0.895($\uparrow$1\%)\\
\OurBase2 &Freezing the bottom 2 layers &70.9M($\downarrow$44\%)&14m06s($\downarrow$18\%)&1h52m48s($\downarrow$27\%)&0.727($\uparrow$1\%)&0.630($\uparrow$3\%)&0.849($\uparrow$1\%)&0.896($\uparrow$1\%)\\
\OurBase3 &Freezing the bottom 3 layers &63.8M($\downarrow$49\%)&13m12s($\downarrow$23\%)&0h39m36s($\downarrow$74\%)&0.724($\uparrow$1\%)&0.626($\uparrow$2\%)&0.842($\uparrow$0\%)&0.895($\uparrow$1\%)\\
\OurBase4 &Freezing the bottom 4 layers &56.7M($\downarrow$55\%)&12m12s($\downarrow$29\%)&0h48m48s($\downarrow$69\%)&0.724($\uparrow$1\%)&0.626($\uparrow$2\%)&0.844($\uparrow$1\%)&0.896($\uparrow$1\%)\\
\OurBase5 &Freezing the bottom 5 layers &49.6M($\downarrow$61\%)&11m42s($\downarrow$32\%)&0h35m06s($\downarrow$77\%)&0.726($\uparrow$1\%)&0.628($\uparrow$3\%)&0.848($\uparrow$1\%)&0.896($\uparrow$1\%)\\
\OurBase6 &Freezing the bottom 6 layers &42.5M($\downarrow$66\%)&10m50s($\downarrow$37\%)&0h32m30s($\downarrow$79\%)&0.727($\uparrow$1\%)&0.629($\uparrow$3\%)&0.848($\uparrow$1\%)&0.896($\uparrow$1\%)\\
\OurBase7 &Freezing the bottom 7 layers &35.4M($\downarrow$72\%)&9m57s($\downarrow$42\%)&0h29m51s($\downarrow$81\%)&0.725($\uparrow$1\%)&0.627($\uparrow$2\%)&0.847($\uparrow$1\%)&0.896($\uparrow$1\%)\\
\OurBase8 &Freezing the bottom 8 layers &28.4M($\downarrow$77\%)&9m04s($\downarrow$47\%)&0h36m16s($\downarrow$77\%)&0.723($\uparrow$0\%)&0.625($\uparrow$2\%)&0.844($\uparrow$1\%)&0.894($\uparrow$1\%)\\
\OurBase9 &Freezing the bottom 9 layers &21.3M($\downarrow$83\%)&8m14s($\downarrow$52\%)&0h24m42s($\downarrow$84\%)&0.718($\downarrow$0\%)&0.620($\uparrow$1\%)&0.838( 0\%)&0.888($\downarrow$0\%)\\
\OurBase10 &Freezing the bottom 10 layers &14.2M($\downarrow$89\%)&7m10s($\downarrow$58\%)&0h21m30s($\downarrow$86\%)&0.710($\downarrow$1\%)&0.612($\downarrow$0\%)&0.829($\downarrow$1\%)&0.882($\downarrow$1\%)\\
\OurBase11 &Freezing the bottom 11 layers &7.1M($\downarrow$94\%)&6m21s($\downarrow$63\%)&0h19m03s($\downarrow$88\%)&0.694($\downarrow$4\%)&0.593($\downarrow$3\%)&0.815($\downarrow$3\%)&0.871($\downarrow$2\%)\\
     \bottomrule
    \end{tabular}
\vspace{-7pt}
\end{table*}

\subsection{Research Question}
\paragraph{RQ3: Are there efficient alternatives to fine-tuning?}
Based on the results of the experimental study, we investigate more efficient alternatives to fine-tune pre-trained code models. Our primary motivation is to freeze the pre-trained parameters of those layers that change only slightly during the fine-tuning of downstream tasks. We propose \textbf{\Our}, which stands for efficient fine-\underline{T}uning of  pr\underline{e}-trained code mode\underline{l}s via  \underline{l}a\underline{y}er freezing. \Our means freezing the pre-trained parameters of the bottom $K$ layers and different $K$ means different variants of our approach.  The $0$-th layer is the embedding layer, and the maximum number of layers of our studied pre-trained code model is 12. If $K$ is set to 12, then \OurBase12 will freeze all parameters and the model will be reduced to the vanilla pre-trained model. Therefore, for the comprehensiveness of the experiments, we vary $K$ from 0 to 11 and conduct extensive experiments on five downstream tasks for these 12 model variants. Next, we introduce the experimental settings, results, and analysis.

\vspace{-5pt}
\subsection{Experimental Settings}
Our experiments are conducted on five diverse downstream tasks including code search, code detection, code summarization, code generation, and line-level code completion. The overview of these tasks is presented in \Tab~\ref{tab:downstream-tasks}. \textit{Code search} is evaluated on the widely-used CodeSeachNet dataset with \python and \ruby and the performance is measured by mean reciprocal rank (MRR) and top-k recall (R@k, k=1,5,10)~\cite{GuoRLFT0ZDSFTDC21,FengGTDFGS0LJZ20,CSNHusain2019}. For \textit{Clone detection}, we experiment on the commonly-used BigCloneBench dataset and use the precision (P), recall (R), and F1-score (F1) as evaluation metrics~\cite{GuoRLFT0ZDSFTDC21,GuoLDW0022}. For \textit{code summarization}, we fine-tune pre-trained models~\cite{GuoRLFT0ZDSFTDC21,GuoLDW0022} on CodeSearchNet dataset with \python and \ruby . The evaluation metrics are sentence-level smoothing BLEU~\cite{bleu02}, Meteor~\cite{meteor05}, Rouge-L~\cite{rouge04} and Cider~\cite{cider15}. \textit{Code generation} is evaluated on the widely-used CONCODE dataset, and the performance is measured by sentence-level smoothing BLEU~\cite{bleu02} and Exact Match accuracy (EM). For \textit{line-level code completion}, We conduct the experiments on the large dataset (named GitHub Java Corpus) in CodeXGLUE. Similar to previous work~\cite{Lu2021}, the performance is measured by EM and Levenshtein edit similarity (Edit sim)~\cite{svyatkovskiy2020intellicode}. The fine-tuning of line-level code completion according to experimental settings in CodeXGLUE~\cite{Lu2021}. Other tasks follow the settings of previous studies~\cite{GuoRLFT0ZDSFTDC21,GuoLDW0022}. We adopt the Adam optimizer with the maximum epoch of 30 and perform early stopping on the validation set. We run the experiments 3 times with random seeds 0,1,2 and display the mean value in the paper. All experiments are conducted on a machine with Tesla A100 GPU. Detailed experimental settings can be found in the online Appendix of the  replication package~\cite{ISSTA2023Repo}.

\vspace{-5pt}
\subsection{Experimental Results}

We conduct experiments with \Our, which freezes the bottom $K$ layers of the pre-trained code model when fine-tuning it on the five downstream tasks. We first present and analyze the experimental results including training parameters, training time cost and performance task by task, and then summarize the findings on the downstream tasks. Due to space limitations, we only present the results of \unixcoder-based \Our and we put results of \graphcodebert-based \Our in the online Appendix~\cite{ISSTA2023Repo}. Conclusions and findings that hold for \unixcoder also generally hold for \graphcodebert. 

\subsubsection{Code search}
\begin{table*}[t]
    \centering
    \setlength{\tabcolsep}{1.6pt}
    \small
    \caption{ Experimental results on clone detection and code summarization. \#Params is short for the number of training parameters. M is short for million. The changing ratios compared to the base model are shown in parentheses. }
   \vspace{-8pt}
    \begin{tabular}{l c cc ccc c  c cc cc}
    \toprule
    \multirow{3}{*}{Model} & \multicolumn{6}{c}{Clone Detection} &\multirow{3}{*}{} &\multicolumn{5}{c}{Code Summarization} \\
      \cmidrule(r){2-7} \cmidrule(r){9-13} 
        & \multirow{2}{*}{\#Params} & \multicolumn{2}{c}{Training time} & \multicolumn{3}{c}{Performance} && \multirow{2}{*}{\#Params} & \multicolumn{2}{c}{Training time} & \multicolumn{2}{c}{Performance} \\
    \cmidrule(r){3-4} \cmidrule(r){5-7}  \cmidrule(r){10-11} \cmidrule(r){12-13} 
    &  &Each epoch &Convergence &Recall &Precision &F1-score  & & &Each epoch &Convergence &BLEU &METEOR  \\ 
\cmidrule(r){1-7} \cmidrule(r){9-13} 
Base &127.1M&20m21s&1h41m45s&0.95&0.95&0.95&&125.93M&22m44s&1h53m40s&19.15&17.26\\
\cmidrule(r){1-7} \cmidrule(r){9-13} 
\OurBase0 &86.2M($\downarrow$32\%)&19m29s($\downarrow$4\%)&1h17m56s($\downarrow$23\%)&0.96($\uparrow$1\%)&0.94($\downarrow$1\%)&0.95( 0\%)&&85.0M($\downarrow$32\%)&21m30s($\downarrow$5\%)&1h47m30s($\downarrow$5\%)&19.19($\uparrow$0\%)&17.32($\uparrow$0\%)\\
\OurBase1 &79.2M($\downarrow$38\%)&18m36s($\downarrow$9\%)&1h14m24s($\downarrow$27\%)&0.95( 0\%)&0.95( 0\%)&0.95( 0\%)&&78.0M($\downarrow$38\%)&20m15s($\downarrow$11\%)&1h41m15s($\downarrow$11\%)&19.21($\uparrow$0\%)&17.33($\uparrow$0\%)\\
\OurBase2 &72.1M($\downarrow$43\%)&17m34s($\downarrow$14\%)&1h10m16s($\downarrow$31\%)&0.95( 0\%)&0.95( 0\%)&0.95( 0\%) &&70.9M($\downarrow$44\%)&19m11s($\downarrow$16\%)&1h35m55s($\downarrow$16\%)&19.17($\uparrow$0\%)&17.30($\uparrow$0\%)\\
\OurBase3 &65.0M($\downarrow$49\%)&15m20s($\downarrow$25\%)&1h01m20s($\downarrow$40\%)&0.95( 0\%)&0.95( 0\%)&0.95( 0\%)&&63.8M($\downarrow$49\%)&17m59s($\downarrow$21\%)&0h53m57s($\downarrow$53\%)&19.16($\uparrow$0\%)&17.26( 0\%)\\
\OurBase4 &57.9M($\downarrow$54\%)&14m28s($\downarrow$29\%)&0h57m52s($\downarrow$43\%)&0.94($\downarrow$1\%)&0.96($\uparrow$1\%)&0.95( 0\%)&&56.7M($\downarrow$55\%)&17m07s($\downarrow$25\%)&0h51m21s($\downarrow$55\%)&19.13($\downarrow$0\%)&17.26( 0\%)\\
\OurBase5 &50.8M($\downarrow$60\%)&13m25s($\downarrow$34\%)&1h07m05s($\downarrow$34\%)&0.96($\uparrow$1\%)&0.94($\downarrow$1\%)&0.95( 0\%) &&49.6M($\downarrow$61\%)&16m18s($\downarrow$28\%)&0h48m54s($\downarrow$57\%)&19.18($\uparrow$0\%)&17.26( 0\%)\\
\OurBase6 &43.7M($\downarrow$66\%)&12m35s($\downarrow$38\%)&0h50m20s($\downarrow$51\%)&0.96($\uparrow$1\%)&0.95( 0\%)&0.95( 0\%)&&42.5M($\downarrow$66\%)&15m10s($\downarrow$33\%)&0h45m30s($\downarrow$60\%)&19.36($\uparrow$1\%)&17.35($\uparrow$1\%)\\
\OurBase7 &36.6M($\downarrow$71\%)&11m44s($\downarrow$42\%)&0h58m40s($\downarrow$42\%)&0.95( 0\%)&0.93($\downarrow$2\%)&0.94($\downarrow$1\%)&&35.4M($\downarrow$72\%)&14m08s($\downarrow$38\%)&0h28m16s($\downarrow$75\%)&19.37($\uparrow$1\%)&17.28($\uparrow$0\%)\\
\OurBase8 &29.5M($\downarrow$77\%)&10m41s($\downarrow$48\%)&0h53m25s($\downarrow$48\%)&0.95( 0\%)&0.94($\downarrow$1\%)&0.95( 0\%)&&28.4M($\downarrow$77\%)&12m59s($\downarrow$43\%)&0h25m58s($\downarrow$77\%)&19.34($\uparrow$1\%)&17.26( 0\%)\\
\OurBase9 &22.4M($\downarrow$82\%)&9m55s($\downarrow$51\%)&0h29m45s($\downarrow$71\%)&0.95( 0\%)&0.92($\downarrow$3\%)&0.93($\downarrow$2\%)&&21.3M($\downarrow$83\%)&11m28s($\downarrow$50\%)&0h22m56s($\downarrow$80\%)&19.18($\uparrow$0\%)&17.22($\downarrow$0\%)\\
\OurBase10 &15.4M($\downarrow$88\%)&8m51s($\downarrow$57\%)&0h35m24s($\downarrow$65\%)&0.97($\uparrow$2\%)&0.92($\downarrow$3\%)&0.94($\downarrow$1\%) &&14.2M($\downarrow$89\%)&10m19s($\downarrow$55\%)&0h10m19s($\downarrow$91\%)&19.11($\downarrow$0\%)&17.18($\downarrow$0\%)\\
\OurBase11 &8.3M($\downarrow$93\%)&8m00s($\downarrow$61\%)&0h32m00s($\downarrow$69\%)&0.96($\uparrow$1\%)&0.92($\downarrow$3\%)&0.94($\downarrow$1\%) &&7.1M($\downarrow$94\%)&09m14s($\downarrow$59\%)&0h09m14s($\downarrow$92\%)&19.10($\downarrow$0\%)&17.20($\downarrow$0\%)\\
     \bottomrule
    \end{tabular}
    \label{tab:freeze_on_clone_summarization}
    \vspace{-10pt}
\end{table*}

We study the performance of 12 \Our variants on code search. The results are shown in \Tab~\ref{tab:freeze_on_search}. We only show the results on \python dataset and put the results on \ruby (which is similar to the results on \python)  in online Appendix~\cite{ISSTA2023Repo}.

In \Tab~\ref{tab:freeze_on_search}, we present the numbers of trained parameters, training time and performance for the base model and 12 model variants. The base model is to fine tune all pre-trained parameters, while the different variants would freeze partial parameters. We report both the training time of each epoch and time til model convergence. 
The changing ratios of different variant models compared to base model are shown in parentheses. From the results of \Tab~\ref{tab:freeze_on_search}, we can find that: 
\begin{itemize}[topsep=0pt,itemsep=0pt,partopsep=0pt,parsep=0pt,leftmargin=10pt]
    \item For all variant models, both the training time cost (especially the convergence time cost) and the training parameters are significantly reduced compared with the base model, while the performance does not change much across the four metrics. Especially, for \OurBase-11 that freezes the bottom 11 layers, the time cost of model convergence and the trained parameters are reduced by 88\% and 94\%, respectively, while the performance only drops by about 3\%.

    \item For \Our ($0 \leq K \leq 8$), they reduce the training parameters by 32\% to 77\%, correspondingly saving about 18\% to 81\% of training time, with the performance increment of 0\% to 3\% across all metrics.

    \item When freezing the bottom 9 layers, there is an 83\% reduction in training parameters and a corresponding 84\% training time saving with a slight change in performance. For example, compared with the base model, the values of MRR and R@10 for \OurBase9 decrease by less than 1\%, R@1 increases by about 1\%, and the R@5 is unchanged.

    \item For \Our ($K \geq 10$), the performance consistently drops across four metrics. However, even freezing the bottom 11 layers, the performance does not drop significantly, while training parameters and corresponding training time are greatly reduced.
\end{itemize}

\begin{tcolorbox}
\textbf{Summary.} In the code search task, the performance of \Our increases for $ 0\leq K \leq 8$, changes slightly for $K=9$,  and decreases lightly for $K\geq10$ compared to the base model.
\end{tcolorbox}

\subsubsection{Clone detection}

We conduct experiments with different \Our variants on clone detection and the results are shown in the left half of  \Tab~\ref{tab:freeze_on_clone_summarization}. As the pre-trained code model adopts 2-layer MLP as the classifier to determine whether two codes are clones or not, the total parameters are about 127.1 million. The performance is evaluated by precision (P), recall (R), and F1-score (F1) and they are in the range of [0, 1]. From  \Tab~\ref{tab:freeze_on_clone_summarization}, we can find that:

\begin{itemize}[topsep=0pt,itemsep=0pt,partopsep=0pt,parsep=0pt,leftmargin=10pt]
    \item For all variant models, training costs and parameters are significantly reduced compared to the base model, while there is no significant change in performance. In particular, when freezing the bottom 11 layers, the convergence time cost and the training parameters are reduced by 88\% and 94\%, respectively, while the performance changes by only 1-3\%.
    
    \item For \Our ($ 0 \leq K \leq 8$), the training parameters are reduced by 32\% to 77\% and correspondingly about 23\% to 51\% training time costs are saved. In addition, the performance is generally stable for F1-score and slightly changes for precision and recall.
    
    \item When ($K \geq 9$), the performance of the different variants consistently decreases in  precision and F1-score but increases in recall. However, even for \OurBase11, all evaluation metrics have high scores (greater than 0.9), while the training parameters and the corresponding time cost are greatly reduced.
\end{itemize}
\begin{tcolorbox}
\textbf{Summary.} In the clone detection task, the performance of \Our is generally stable for $ 0 \leq K \leq 8$ but lightly changes for $K \geq 9$ compared to the base model.
\end{tcolorbox}

\subsubsection{Code summarization}

We conduct the experiments with different \Our on code summarization and the results are shown in the right half of  \Tab~\ref{tab:freeze_on_clone_summarization}. The results of Rouge-L and Cider on \python and all experimental results on \ruby are put in online Appendix~\cite{ISSTA2023Repo} due to space limit. The reported metrics including BLEU and METEOR are in the range of [0, 100]. From \Tab~\ref{tab:freeze_on_clone_summarization}, we can find that:
\begin{itemize}[topsep=0pt,itemsep=0pt,partopsep=0pt,parsep=0pt,leftmargin=10pt]
    \item For all variant models, both training time costs and parameters are significantly reduced compared to the base model, while the performance does not change much for all  metrics.
    
    \item For \Our ($0 \leq K \leq 5$), they reduce the training parameters by 32\% to 61\%, correspondingly saving about 5\% to 57\% of training time, with stable performance. In particular, the performance change is less than 1\% for all evaluation metrics. 
    
    \item For \Our ($6 \leq K \leq 9$), training parameters are reduced by 66\% to 83\%, corresponding to 60\% to 80\% saving in training time with a generally slight increase in performance. 

    \item When \Our ($K \geq 10$), the performance slightly drops in terms of four metrics. However, even freezing the bottom 11 layers, the performance does not drop significantly, while both the training parameters and the corresponding time cost are extremely reduced.
\end{itemize}

\begin{tcolorbox}
\textbf{Summary.} On the code summarization task, the performance of \Our is stable for ($ 0 \leq K \leq 5$),  slightly increases  for $6 \leq K \leq 9$, but lightly decreases when $K \geq 10$ compared to the base model.
\end{tcolorbox}

\subsubsection{Code generation}

\begin{table*}[ht]
    \centering
    \setlength{\tabcolsep}{1.6pt}
    \small
    \caption{Experimental results on code generation and line-level code completion. \#Params is short for the number of training parameters. M is short for million. The changing ratios compared to the base model are shown in parentheses.}
    \vspace{-5pt}
     \vspace{\tabmargin}
    \label{tab:freeze_on_code_gen_completion}
    \begin{tabular}{c ccccc c ccccc}
    \toprule
    \multirow{3}{*}{Model} & \multicolumn{5}{c}{Code Generation} &\multirow{3}{*}{} &\multicolumn{5}{c}{Line-Level Code Completion} \\
      \cmidrule(r){2-6} \cmidrule(r){8-12} 
        & \multirow{2}{*}{\#Params} & \multicolumn{2}{c}{Training time} & \multicolumn{2}{c}{Performance} && \multirow{2}{*}{\#Params} & \multicolumn{2}{c}{Training time} & \multicolumn{2}{c}{Performance} \\
    \cmidrule(r){3-4} \cmidrule(r){5-6}  \cmidrule(r){9-10} \cmidrule(r){11-12} 
    &  &Each epoch &Convergence  &BLEU &EM   &&  &Each epoch &Convergence  &Edit sim &EM \\ 
\cmidrule(r){1-6} \cmidrule(r){8-12} 
Base &125.93M&12m25s&4h08m20s&33.82&17.4&&125.93M&04m12s&0h37m48s&51.92&20.40\\
\cmidrule(r){1-6} \cmidrule(r){8-12} 
\OurBase0 &85.0M($\downarrow$32\%)&11m43s($\downarrow$6\%)&3h42m37s($\downarrow$10\%)&33.88($\uparrow$0\%)&18.1($\uparrow$4\%)&&85.0M($\downarrow$32\%)&03m59s($\downarrow$5\%)&0h31m52s($\downarrow$16\%)&52.58($\uparrow$1\%)&21.07($\uparrow$3\%)\\
\OurBase1 &78.0M($\downarrow$38\%)&11m07s($\downarrow$10\%)&3h20m06s($\downarrow$19\%)&34.43($\uparrow$2\%)&17.9($\uparrow$3\%)&&78.0M($\downarrow$38\%)&03m47s($\downarrow$10\%)&0h18m55s($\downarrow$50\%)&52.35($\uparrow$1\%)&20.87($\uparrow$2\%)\\
\OurBase2 &70.9M($\downarrow$44\%)&10m32s($\downarrow$15\%)&2h06m24s($\downarrow$49\%)&33.85($\uparrow$0\%)&19.1($\uparrow$10\%)&&70.9M($\downarrow$44\%)&03m38s($\downarrow$13\%)&0h32m42s($\downarrow$13\%)&52.31($\uparrow$1\%)&20.93($\uparrow$3\%)\\
\OurBase3 &63.8M($\downarrow$49\%)&09m52s($\downarrow$21\%)&2h08m16s($\downarrow$48\%)&34.24($\uparrow$1\%)&19.0($\uparrow$9\%)& &63.8M($\downarrow$49\%)&03m29s($\downarrow$17\%)&0h20m54s($\downarrow$45\%)&51.81($\downarrow$0\%)&20.73($\uparrow$2\%)\\
\OurBase4 &56.7M($\downarrow$55\%)&09m15s($\downarrow$26\%)&1h51m00s($\downarrow$55\%)&34.02($\uparrow$1\%)&18.6($\uparrow$7\%)&&56.7M($\downarrow$55\%)&03m18s($\downarrow$21\%)&0h26m24s($\downarrow$30\%)&51.62($\downarrow$1\%)&20.27($\downarrow$1\%)\\
\OurBase5 &49.6M($\downarrow$61\%)&08m44s($\downarrow$30\%)&1h53m32s($\downarrow$54\%)&34.36($\uparrow$2\%)&18.6($\uparrow$7\%)&&49.6M($\downarrow$61\%)&03m07s($\downarrow$26\%)&0h09m21s($\downarrow$75\%)&51.66($\downarrow$0\%)&20.40(0\%)\\
\OurBase6 &42.5M($\downarrow$66\%)&08m13s($\downarrow$34\%)&1h46m49s($\downarrow$57\%)&32.90($\downarrow$3\%)&18.1($\uparrow$4\%)&&42.5M($\downarrow$66\%)&02m57s($\downarrow$30\%)&0h11m48s($\downarrow$69\%)&51.36($\downarrow$1\%)&20.27($\downarrow$1\%)\\
\OurBase7 &35.4M($\downarrow$72\%)&07m41s($\downarrow$38\%)&1h39m53s($\downarrow$60\%)&32.92($\downarrow$3\%)&18.0($\uparrow$3\%)&&35.4M($\downarrow$72\%)&02m58s($\downarrow$29\%)&0h14m50s($\downarrow$63\%)&51.32($\downarrow$1\%)&20.40( 0\%)\\
\OurBase8 &28.4M($\downarrow$77\%)&07m08s($\downarrow$43\%)&1h32m44s($\downarrow$63\%)&32.12($\downarrow$5\%)&17.4( 0\%)&&28.4M($\downarrow$77\%)&02m59s($\downarrow$29\%)&0h14m55s($\downarrow$65\%)&50.95($\downarrow$2\%)&20.27($\downarrow$1\%)\\
\OurBase9 &21.3M($\downarrow$83\%)&06m42s($\downarrow$46\%)&1h27m06s($\downarrow$65\%)&31.33($\downarrow$7\%)&18.1($\uparrow$4\%)&&21.3M($\downarrow$83\%)&02m58s($\downarrow$29\%)&0h14m50s($\downarrow$67\%)&51.13($\downarrow$2\%)&20.73($\uparrow$2\%)\\
\OurBase10 &14.2M($\downarrow$89\%)&06m03s($\downarrow$51\%)&1h12m36s($\downarrow$71\%)&30.43($\downarrow$10\%)&16.6($\downarrow$5\%)&&14.2M($\downarrow$89\%)&02m57s($\downarrow$30\%)&0h11m48s($\downarrow$76\%)&50.70($\downarrow$2\%)&20.53($\uparrow$1\%)\\
\OurBase11 &7.1M($\downarrow$94\%)&05m22s($\downarrow$57\%)&1h09m46s($\downarrow$72\%)&27.71($\downarrow$18\%)&14.3($\downarrow$18\%)& &7.1M($\downarrow$94\%)&02m06s($\downarrow$50\%)&0h06m18s($\downarrow$83\%)&49.49($\downarrow$5\%)&19.67($\downarrow$4\%)\\
     \bottomrule
    \end{tabular}
\vspace{-10pt}
\end{table*}

We conduct experiments with different \Our on code generation and the results are shown in the left half of \Tab~\ref{tab:freeze_on_code_gen_completion}. The evaluation metrics including BLEU and EM are in the range of [0, 100]. From \Tab~\ref{tab:freeze_on_code_gen_completion}, we can find that:

\begin{itemize}[topsep=0pt,itemsep=0pt,partopsep=0pt,parsep=0pt,leftmargin=10pt]
    \item For all variant models, both training time costs and parameters are significantly reduced compared to the base model, while the performance does not change much for all metrics except for \Our{}$(K \geq 8)$.
    
    \item For \Our ($ 0 \leq K \leq 5$), they reduce the training parameters by 32\% to 61\%, correspondingly saving about 10\% to 55\% of training time, and the performance is slightly improved. Specifically, the BLEU scores are increased by 0\% to 2\%, and the EM scores are increased by 3\% to 10\%.
    
    \item For \Our ($6 \leq K \leq 9$), training parameters are reduced by 66\% to 83\%, corresponding to 57\% to 65\% saving in training time. The performance of these variants significantly drops under BLEU but generally increases under EM. This is because BLEU combines the n-gram precision (n=1,2,3,4) between the generated code snippet and the ground truth for one sample, while EM is 1 if they are exactly the same, 0 otherwise. However, for the entire set, only about 18\% of the code snippets can be generated exactly the same as ground truth. The other ~82\% samples also affect the final result of BLEU scores. Therefore, the performance changes for BLEU and EM behave differently.
    
    \item When \Our ($  K \geq 10$), training parameters and the corresponding time cost are greatly reduced, and the performance of variants also significantly drops on both metrics. 
\end{itemize}
\begin{tcolorbox}
\textbf{Summary.} On the code generation task, the performance of \Our is slightly improved for \Our ($ 0 \leq K \leq 5$), obviously changes for $ 6 \leq K \leq 9$, and significantly drops when $K \geq 10$ compared to the base model.
\end{tcolorbox}

\subsubsection{Line-level code completion}
We conduct the experiments with all \Our on line-level code completion and the results are shown in the right half of  \Tab~\ref{tab:freeze_on_code_gen_completion}. The evaluated metrics including Edit Sim and  EM are in the range of [0, 100]. From \Tab~\ref{tab:freeze_on_code_gen_completion}, we can find that:

\begin{itemize}[topsep=0pt,itemsep=0pt,partopsep=0pt,parsep=0pt,leftmargin=10pt]
    \item For all variant models, both the training time cost and parameters are greatly reduced compared to the base model, while the performance does not change significantly except the \OurBase11.
    
    \item For \Our ($ 0 \leq K \leq 7$), they reduce the training parameters by 32\% to 72\%, correspondingly saving about 13\% to 75\% of training time. In addition, the performance of \Our ($ 0 \leq K \leq 3$) is lightly improved and the performance of \Our ($ 4 \leq K \leq 7$) slightly drops.
    
    \item For \Our ($  K \geq 8$), training parameters and corresponding time cost are hugely reduced. The performance of variants generally drops for two metrics. Especially, when $K=11$, the performance significantly drops.  Therefore, the pre-trained parameters of the last layers need to be fine-tuned to learn to predict the next line of code.
\end{itemize}

\begin{tcolorbox}
\textbf{Summary.} On code completion task, the performance of \Our is slightly improved for $ (0 \leq K \leq 10$ and significantly drops when $K=11$ compared to the base model.
\end{tcolorbox}

\subsubsection{Findings across all downstream tasks}
After analyzing the experimental results task by task,  we summarize the general findings across various tasks as follows.

\begin{itemize}[topsep=0pt,itemsep=0pt,partopsep=0pt,parsep=0pt,leftmargin=10pt]

    \item For  all \Our, both the training time cost (especially the convergence time cost) and the training parameters are significantly reduced compared to the base model. The performance does not change much, except for \OurBase10 and \OurBase11 on code generation and  \OurBase11 on code completion.
    
    \item When $0 \leq K \leq 5$, \Our generally reduces the training parameters by 30\% to 65\%, correspondingly saving about 10\% to 70\% of training time, with the performance generally increasing varying from 1\% to 4\% in terms for various downstream tasks. Therefore, \OurBase5 is usually the best choice among the variants which reduce resource consumption and performance improvement.
    
    \item When $6 \leq K \leq 9$, \Our generally reduces the training parameters by  65\% to 80\% , correspondingly saving about  50\% to 80\% of training time, with the performance generally marginally changes for various downstream tasks. Specifically, the performance of code search and code generation varies from 1\% to 5\%, clone detection and code summarization varies from 0\% to 1\%, and code completion varies from 0\% to 3\%. Therefore, \OurBase9 is usually the best choice among the variants which reduce resource consumption with a small change in performance.
    
    \item When $ K \geq 10$,  both training parameters and time cost are greatly reduced. The performance of \Our drops but not significantly except for code generation and code completion. 
    
\end{itemize}

The performance of models increases despite freezing some layers in the above experiments. One possible explanation is that the reduction in the number of parameters alleviates overfitting, allowing parameters fine-tuned on the training set to better generalize to the testing set. However, the exact mechanism behind this phenomenon is still not well understood and requires further investigations, such as studying more datasets with different distributions between training and testing sets, to verify our conjecture.

\begin{tcolorbox}
\textbf{Summary.} For \Our across various downstream tasks, both the training parameters and the time cost are extremely reduced compared to the base model. In general, the performance of \Our increases by 1\% to 4\% for $0 \leq K \leq 5$, slight changes for $6 \leq K \leq 9$, and obviously drops when $K\geq 10$ compared to the base model.
\end{tcolorbox}

\vspace{-5pt}
\section{Discussions and Threats to Validity}
\label{sec:discuss_threats}

\subsection{Importance of Reducing Fine-tuning Time Costs and the Advantages of \Our{}.}
It is important to reduce fine-tuning costs especially time costs because (1) the pre-training with fine-tuning paradigm shows increasing adoption in many software engineering tasks~\cite{AhmadCRC21,Lu2021,FengGTDFGS0LJZ20,GuoLDW0022}. Reducing time cost is important as it typically takes much time to fine-tune a model, especially on larger datasets. Reducing fine-tuning time costs can also improve the development efficiency of pre-trained code models, especially when developers need to meet product launching deadlines, thereby saving costs and reducing unexpected losses. (2) GPUs are usually expensive computing resources. We can save GPU resources with reduced time costs. Moreover.  Training deep-learning-based models, including fine-tuning pre-trained models, can emit over 600K tons of CO2 each year~\cite{StrubellGM20}. By reducing time costs, carbon emissions can be reduced.

In fact, many approaches~\cite{wang2020k,lu2020twinbert,jiang2022transferability,JiaoYSJCL0L20,HoulsbyGJMLGAG19,wang2020minilm} have been proposed to save fine-tuning time costs. Among them, distillation techniques~\cite{lu2020twinbert,JiaoYSJCL0L20,wang2020minilm} are promising and popular. Specifically, they propose different approaches to compress a large pre-trained model into a smaller model, and fine-tune the smaller model to perform downstream tasks. However, these techniques need to carefully design and tune the architectures of small models and loss functions to distill knowledge from big models. Therefore, generally the process of distillation requires more manual design and is more laborious. Our approach~\Our{} is simple and can directly decrease the training cost via layer freezing. 

\vspace{-10pt}
\subsection{How to help better fine-tuning in the future}
In this paper, we have not concluded a ``one-method-to-rule-them-all'' suggestion for different tasks. However, our study of RQ2 and RQ3 show that \OurBase7, which freezes the bottom 7 layers, achieves a good trade-off between efficiency and performance. Specifically, (1) from~\Fig~\ref{fig:rsa_search_py}, we can see that the representations of the bottom 7 layers between pre-trained and fine-tuned models are very similar; (2) from ~\Tabs~\ref{tab:freeze_on_search},~\ref{tab:freeze_on_clone_summarization} and~\ref{tab:freeze_on_code_gen_completion}, we can see that \OurBase-7 significantly reduces the training time and parameters while maintaining similar performance for all downstream tasks. Therefore, it is recommended to use~\OurBase-7 in practice.  In future work, we aim to investigate automated layer freezing strategies to use layer-wise representations more efficiently. For example, we plan to design algorithms to automatically select features from different layers and aggregate them to perform different downstream tasks.

\vspace{-10pt}
\subsection{Threats to Validity}
We have identified the following threats to our study:

\emph{Program Languages.} We conduct experiments on five programming languages (\python, \java, \ruby, C, and C++). Although in principle, our studied models are not specifically designed for certain languages, models may perform differently on different programming languages. Therefore, more experiments are needed to confirm the generality of our findings and conclusions. In the future, we will extend our study to more programming languages.

\emph{Evaluation Datasets.} We conduct the experiments on widely-used datasets. Besides, there are other datasets for each downstream task. They are different in some aspects such as construction methods and corpus sizes. Model may perform differently on different datasets. Thus, we will conduct experiments on more datasets to confirm the generality of our findings and conclusions.

\emph{Evaluation Metrics.} We use as many commonly-used metrics as possible to evaluate model performance in this study. However, these metrics may have their inherent limitations. For example, BLEU and METEOR are textual similarity-based metrics and cannot measure the semantic similarity of two sentences. In the future, we will use more metrics and human evaluation to confirm the findings and conclusions in this study.

\emph{Pre-trained Code Models.} Due to computational resource constraints, we focus on the state-of-art pre-trained code model \unixcoder and \graphcodebert in this study. Other pre-trained code models such as CodeGPT~\cite{Lu2021} and CodeT5~\cite{codet5wang} are yet to be studied.
 
 \section{Related Work}
 \label{sec:related_work}
\subsection{Probing Pre-trained Models}
In the natural language processing community, many studies~\cite{MerchantRPT20,ChrupalaA19,AbnarBCZ19,TenneyDP19,TenneyXCWPMKDBD19} have investigated how pre-trained language models understand natural language and what happens when fine-tuning them. They are typically divided into two categories. The first employs probing techniques to study what linguistic properties are captured by pre-trained language models~\cite{TenneyDP19,TenneyXCWPMKDBD19}. The second is representational similarity analysis~\cite{AbnarBCZ19,ChrupalaA19,MerchantRPT20}. It is a task-agnostic analysis and is used to measure the similarity between two different representational spaces. However, in software engineering field, few studies explore what happens to pre-trained code models \emph{during the fine-tuning process}. Most related studies~\cite{FengGTDFGS0LJZ20,WanZZSXJ22,troshin2022probing,karmakar2021pre,hernandez2022ast} aim to understand what pre-trained code models know about source code. For example, Wan et al.~\cite{WanZZSXJ22} conduct a structural analysis to demonstrate that the pre-trained models are aware of syntactic structure. López et al~\cite{hernandez2022ast} recover ASTs from hidden representations of pre-trained language models. Karmakar et al~\cite{karmakar2021pre} also propose some new probing tasks to investigate what pre-trained code models know about code. Inspired by the compilation process and static analysis, we first propose four probing tasks involving the lexical, syntactical, semantic, and structural code
properties. Then we investigate what code properties are encoded in layer-wise pre-trained representations and what happens to these representations during fine-tuning.
\vspace{-5pt}
\subsection{Accelerating the Fine-tuning Process}
There are many studies on accelerating fine-tuning process~\cite{wang2020k,lu2020twinbert,jiang2022transferability,JiaoYSJCL0L20,HoulsbyGJMLGAG19,wang2020minilm}. These studies can be roughly categorized into two categories. The first is to use the knowledge distillation technique to compress large-scale pre-trained language models~\cite{lu2020twinbert,JiaoYSJCL0L20,wang2020minilm}. For example, Jiao et al.~\cite{JiaoYSJCL0L20} propose TinyBERT to distill BERT and only use about 28\% parameter for natural language understanding. The second is the adapter-based fine-tuning approach~\cite{wang2020k,HoulsbyGJMLGAG19}, where adapters are new trainable modules added between layers of pre-trained models. For example, Houlsby et al.l~\cite{HoulsbyGJMLGAG19} design some adapters with two orders of magnitude fewer parameters to fine-tune compared to full models and achieve similar performance with fine-tuning all parameters of the pre-trained mode. In addition, there are some studies on efficient neural network training from scratch with layer freezing~\cite{jiang2020unified,WangKGfreeze2022}. For example, Wang et al.~\cite{WangKGfreeze2022} leverage the knowledge from a reference model to accurately evaluate individual layers’ training plasticity, freeze the converged ones and unfreeze the frozen layers to continue training. Our study could motivate researchers to come up with more efficient fine-tuning approaches.                       

\vspace{-2pt}
\section{Conclusion}
\label{sec:conclusion}
In this paper, we firstly conduct extensive experimental study to explore what happens to layer-wise code knowledge and pre-trained representations during fine-tuning. We then propose efficient alternatives to fine-tune the large pre-trained code model based on the above findings. Our experimental study shows that the lexical, syntactic, and \structural properties of source code are mainly captured in the lower, intermediate, and higher layers, respectively, while the semantic property spans across the entire model. The basic code properties captured by lower and intermediate layers are still preserved during fine-tuning. Furthermore, we find that only the representations of the top two layers change the most during fine-tuning for various downstream tasks. Based on the above findings, we propose \Our that efficiently fine-tunes pre-trained code models via selective layer freezing. The extensive experiments on five various downstream tasks demonstrate that both training parameters and time costs can be greatly reduced, while performance is similar or even better.

Furthermore, our experimental study shows many useful findings and promising directions for efficient fine-tuning of pre-trained code models. For example,  we find that pre-trained code models encode syntactic properties into intermediate layers. Therefore, when the downstream tasks are syntax-aware such as code generation, we could design some modules to make use of the contextual representations of intermediate layers. 
In the future, we will continue to explore various promising directions and propose more  efficient fine-tuning approaches.

Replication package including source code, datasets, and online Appendix is available at: \url{https://github.com/DeepSoftwareAnalytics/Telly}.
\vspace{-2pt}
\section*{Acknowledgement}
We thank reviewers for their valuable comments on this work. This research was supported by National Key R\&D Program of China (No. 2017YFA0700800) and Fundamental Research Funds for the Central Universities under Grant xtr072022001.

 \bibliographystyle{ACM-Reference-Format}
\bibliography{ref}
\end{document}